\documentclass[a4paper, 12pt]{article}
        \title{\sf Construction of 2nd stage shape invraiant potentials}
        \author{S. Sree Ranjani \\
      Faculty of Science of Technology,\\ ICFAI foundation for Higher
Education,\\
(Declared as deemed-to-be University u/s 3 of the UGC Act 1956),\\
Dontanapally, Hyderabad, India, 501203.\\[3mm]
 and \\[3mm]
A. K. Kapoor\\
School of Basic Sciences,\\ I.I.T. Bhubaneswar, Main Campus Argul,\\
Bhubaneswar 752050, INDIA
}
\usepackage{amsmath}
\usepackage{amssymb}
\newcommand{\dd}[2][]{\frac{d #1}{d#2}}
\newcommand{\DD}[2][]{\frac{d^2 #1}{d#2^2}}
\begin{document}
              %Comment out/Modify Whatever is is not needed
              \baselineskip=16pt
              \maketitle
              \begin{abstract}
                  We introduce concept of next generation shape invariance and
                  show that the process of shape invariant extension can be
                  continued indefinitely.
              \end{abstract}

%%%-------------------------------------------------------------------------
\subsubsection*{Introduction}
Suppose we have a shape invariant potential \(V(x)\) with
superpotential given by \(W(x)\), prepotential by \(\Omega(x)\),
the partner potentials \(V_\pm(x)\) defined by
\begin{equation}\label{EQ01}
  V_\pm = W^2 - W^\prime(x)
\end{equation}
differ from \(V(x)\) by a constant. Shape invariance means that  \(V_-(x)\) and
\(V_+(x)\) have the same form and coincide if parameters appearing in
\(W(x\) are redefined.

In an earlier papers we have demonstrated that all the shape invariant
potentials could be constructed by assuming an ansatz \(W(x) = \lambda F(x)\)
and solving the constraint implied by shape invariance [1,2]. In addition a
process, of arriving at the shape invariant  potentials related
to the exceptional orthogonal polynomials,  was given within the framework of
quantum Hamilton Jacobi (QHJ) formalism. This naturally raises the question if
the process can be repeated  to arrive at potentials related to next generation
of exceptional polynomials. This question has been addressed in a recent
article by Sree Ranjani within the frame work of QHJ [3]. She has explicitly
constructed the second generation potentials and the corresponding polynomials.

Here we give a slightly different approach to the above mentioned problem.
It will be seen in that this approach is much simpler and leads to a set of new
results on shape invariant potentials.

If introduce a function \(f(x)\) defined by
\[f(x) =\exp\Big(-\int W(x) dx\Big),\]
the Hamiltonians corresponding to the potentials \(V_\pm(x)\) in \eqref{EQ01}
can be written as
\begin{equation}\label{EQ02}
  H_+ = - \frac{1}{f}\dd{x} f^2 \dd{x} \frac{1}{f}, \qquad
  H_- = - f\dd{x}\frac{1}{f^2}\dd{x} f.
\end{equation}
{\it We will assume that the partners \(H_\pm\) have the shape invariance
property and that the solutions of the corresponding eigenvalue equations
are known.}

\subsubsection*{Generalizing Shape Invariance}
Given a differential operator of the form  \(H_\pm\) in \eqref{EQ02}we
introduce  a differential operators
\begin{eqnarray}
  L_1 = - \frac{1}{g}\dd{x} g \dd{x} ; \qquad L_2 = - g\dd{x} \frac{1}{g}\dd{x}.
\end{eqnarray}
and define
\begin{equation}\label{EQ04}
  L \equiv L_1 + \widetilde{V}(x)  = - \frac{1}{g}\dd{x} g \dd{x} +
\widetilde{V}(x)
\end{equation}
It can be easily seen that, for the choice  \(g=f^2\), the differential
operators \(L_1,L_2\) are related to \(H_\pm\).
\begin{equation}
  L_1={f}^{-1} H_+ f ,\qquad\qquad L_2 =  f H_- {f}^{-1}.
\end{equation}

The additional term  \(\widetilde{V}(x)\) in Eqref{EQ04} in the
potential \(V(x)\) will called a next
generation term in the potential.
One the aims of this papers is to use a generalized shape
invariance requirement and to determine restrictions on
the next generation term \(\widetilde{V}(x)\).

We  define a concept of {\bf gen-next shape
invariance}  for \(\widetilde{V}(x)\) in a manner close to the  shape
invariance property as it is understood in the current literature.  For  this
purpose we follow our paper-I.

Given a Schrodinger differential operator \(L\), of the form \eqref{EQ04},
we will define partner potentials and gen-next shape invariance as follows.
We first set up the eigenvalue equation for
\begin{equation}
  L \psi = E \psi.
\end{equation}
We write \(\psi(x) = \exp(\Omega(x))\), where \(\Omega(x)\) obeys the QHJ
equation for \(L\).
\begin{equation}
  \Big(\dd[\Omega]{x}\Big)^2 + \frac{1}{g} \dd{x}\Big(g\dd[\Omega(x)]{x}\Big) -
E  = 0.
\end{equation}
The above form suggests that we introduce next-gen partner potentials
\(\widetilde{V}_\pm(x)\) by
\begin{equation}
  \widetilde{V}_\pm(x) = W^2 \pm \frac{1}{g}\dd{x}(gW)
\end{equation}
where  \(W(x)= \dd[\Omega]{x}\).
In general the function \(W(x)\), to be  by next-gen super potential,
will depend on some parameters. In the simplest
case of translational shape invariance,  there is only one such parameter.
Anticipating the results, we assume that the next-gen superpotential potential
\(W(x)\)
can be chosen to depend on one parameter. We use \(\lambda, \mu\) etc. to denote
different  values of this parameter. The shape invariance requirement,
\( V_+(x,\lambda) = V_-(x,\mu) + \text{const}\), then take the form
\begin{equation}
  W^2(x,\lambda) + \frac{1}{g(x)} \dd{x}\Big( g(x) W(x,\lambda) \Big)
  =   W^2(x,\mu) - \frac{1}{g(x)} \dd{x}\Big( g(x) W(x,\mu) \Big) +
\text{constant}
\end{equation}
As before, following  our paper [1], we make an ansatz
\begin{equation}
  W(x) = \lambda F(x).
\end{equation}
and rewrite the resulting equation in the form
\begin{equation}
 F^2 + \frac{1}{(\mu-\lambda)}\frac{1}{g} \dd[(gF)]{x}= \text{constant}.
\end{equation}
Next change the variable from \(x\) to \(\xi =\alpha x\), where \(\alpha =
\mu-\lambda\). Using notation \(\bar{g}(\xi)= g(\xi/\alpha)\) we get
\begin{equation}\label{EQ11}
  F^2(\xi) + \frac{1}{\bar{g}}\dd{\xi}\Big(\bar{g} F \Big) =  K,
\end{equation}
where \(K\) is a constant. Next substituting \(F= \psi^\prime/\psi\), the
function \(\psi\) satisfies  the equation
\begin{equation}
  \DD[\psi]{\xi} + \frac{1}{\bar{g}}\dd[\bar{g}]{\xi} \, \dd[\psi]{\xi} -  K
     \psi = 0.
\end{equation}
This equation is can be rearranged as
\begin{equation}
 - \Big(\frac{1}{\bar{g}}\dd{\xi} \bar{g} \dd{\xi} + K  \Big) \psi=0.
\psi + K \psi=0.
\end{equation}
Substituting  \(g= f^2; \psi= \frac{1}{f}\phi\), we get
\begin{equation}
 \Big( \frac{1}{f}\dd{\xi} f^2 \dd{\xi} \frac{1}{f} + K \Big) \phi(\xi) =0
\end{equation}
The solutions of this equation are known and can be used to construct solution
for the function \(F(x)\) and hence the next generation of shape invariant
potentials can be determined. It  is obvious that the process outlined here can
be carried out indefinitely.
\subsubsection*{Remarks}
\begin{itemize}
\item It should be noted that in the second stage we will get a two parameter
shape invariant potential. In general we will have translational invariance in
\(2^n\) parameters in the \(n^\text{th}\) stage.
\item Instead of starting from \(L_1\) we could have started with
\(L_2\) see \eqref{EQ04}. This will lead to two other potentials. Thus we will
have 'four partners' in the second stage and \(2^n\) partners after the
\(n^\text{th}\) stage. The multiparameter shape invariant potentials have been
found for the first time in this work.

\item The EOP and related potentials discussed in the literature correspond to
special values of the parameter \(\lambda\) appearing in higher generation
potentials.
\item  {\it Iso-spectral deformation route:}
Suppose  we have potentials \(V_\pm\) specified by superpotential \(W(x)\) and
that  we wish to carryout a shape invariant extension, following
our earlier paper.  We would then write
\(\overline{W}(x) = W(x)+\chi(x)\) and demand that \(V_+\) (or \(V_-)\)
\(W(x)\) and \(\overline{W}(x)\) be equal up to a constant, we would get
\begin{equation}
  W^2(x) + \dd[W]{x}  =   \overline{W}^2(x) + \dd[\overline{W}]{x}
\end{equation}
This route to obtain a new shape invariant potential leads to the following
constraint on \(\chi\)
\begin{eqnarray}
  \chi^2 \pm 2W(x) \chi + \chi^\prime +K=0 \\
   \chi^2 \pm  2\chi\Big(\frac{1}{f} \dd{x}f(x)\Big)   + \chi^\prime +K=0
\end{eqnarray}
because \(W=f^\prime/f\). One of these equations coincides with
\eqref{EQ11}  if we choose  the sign,
define the argument suitably and remember that \(\bar{g} = f^2\). Thus the
process of going to the next generation of shape invariance described
here contains the isospectral deformation process used in our paper [1].
\end{itemize}

\subsubsection*{Concluding reamrks} In this paper we have demonstrated for any
shape invariant potential, for which the nergye eigenfucntions and eigenvalues
are known, we  can construct the next generation shape invaraint potential and
the process can be repeated indefinitely. Physically only potentials which
are free of singularities are of interest. To decide if the next generation
potential so obtained  is free of singularity or not  requires a separate
study.

\subsubsection*{References}
\begin{itemize}
\item[{[1]}] S. Sree Ranjani, R. Sandhya and A. K. Kapoor, ``Shape invariant
rational extensions and potentials related to exceptional polynomials"\\
International Journal of Modern Physics A {\bf 30} (2015) 1550146.
\item[{[2]}]  R. Sandhya, S. Sree Ranjani and A. K. Kapoor ''Shape invariant
potentials in higher dimensions''\\
Annals of Physics 359 (2015) 125–135.
\item[{[3]}] S.S. Ranjani , ``QHJ route to multi-indexed exceptional Laguerre
polynomials and corresponding rational potentials'', arrXiv.org :1705.07682.
\end{itemize}
\end{document}